\documentclass[twocolumn,english,showpacs,preprintnumbers,amsmath,amssymb,floatfix]{revtex4}
\usepackage[T1]{fontenc}
\usepackage[latin9]{inputenc}
\usepackage{color}
\usepackage{array}
\usepackage{amstext}
\usepackage{graphicx}
\usepackage{esint}

\makeatletter

\@ifundefined{textcolor}{}
{%
 \definecolor{BLACK}{gray}{0}
 \definecolor{WHITE}{gray}{1}
 \definecolor{RED}{rgb}{1,0,0}
 \definecolor{GREEN}{rgb}{0,1,0}
 \definecolor{BLUE}{rgb}{0,0,1}
 \definecolor{CYAN}{cmyk}{1,0,0,0}
 \definecolor{MAGENTA}{cmyk}{0,1,0,0}
 \definecolor{YELLOW}{cmyk}{0,0,1,0}
 }

\@ifundefined{definecolor}
 {\usepackage{color}}{}
\@ifundefined{definecolor}
 {\usepackage{color}}{}
\makeatother

\makeatother

\usepackage{babel}

\begin{document}

\title{Spin-dependent fragmentation functions of Gluon splitting into heavy quarkonia
considering three different scenarios}

\author{S. Mohammad Moosavi Nejad$^{a,b}$}
\email{mmoosavi@yazd.ac.ir}

\author{Mahdi Delpasand$^a$}

\affiliation{$^{(a)}$Faculty of Physics, Yazd University, P.O. Box
89195-741, Yazd, Iran}

\affiliation{$^{(b)}$School of Particles and Accelerators,
Institute for Research in Fundamental Sciences (IPM), P.O.Box
19395-5531, Tehran, Iran}

\date{\today}

\begin{abstract}
Heavy quarkonium production is a powerful implement to study the 
strong interaction dynamics and QCD theory.
Fragmentation is the dominant production mechanism for heavy quarkonia with large transverse momentum.
With the large heavy quark mass, the relative motion
of the heavy quark pair inside a heavy quarkonium is effectively nonrelativistic and
it is also well-known that their fragmentation functions  can be calculated in
the perturbative QCD framework. Here, we analytically calculate the process-independent  
fragmentation functions for a gluon to split into the spin-singlet 
and -triplet $S$-wave heavy quarkonia 
using three different scenarios.  We will show that the fragmentation 
probability of the gluon into the spin-triplet bound-state  is the biggest one.

\end{abstract}


\maketitle

\section{Introduction}
\label{sec:intro}

Heavy quarkonia are the simplest particles when the strong interactions are concerned.
Studying their production mechanism 
is important for understanding QCD and the strong interaction dynamics.
In theory, it is well-known that the dominant mechanism to produce a heavy quarkonium at large 
transverse momentum is fragmentation \cite{Braaten:1993rw} so this
mechanism is described by the fragmentation functions (FFs). In fact, the FF shows hadron production probability 
from the initial high-energy parton. The FFs are
universal quantities so that are independent of the initial parton production processes.
The particular importance of FFs is for model-independent predictions of the cross sections at the Large Hadron Collider (LHC).
Basically, the fragmentation is related to the low-energy part of the hadron production
process and forms the nonperturbative aspect of QCD so, in principle, should be extracted from experimental
data. 
In practice, it is hard to extract so many initial FFs from data directly.\\
In Ref.~\cite{Soleymaninia:2013cxa}, using the phenomenological approach 
which is based on the data analysis
we propounded a new functional form of pion and kaon FFs up to NLO,
obtained through a global fit to the data from the single-inclusive annihilation $e^-e^+\rightarrow hX$ and also from 
the semi-inclusive DIS asymmetry data from HERMES and COMPASS.\\
Since heavy quarkonia have simple internal structures  it is well-known that the perturbative QCD approximations to their
FFs  are well-defined in the nonrelativistic QCD factorization framework 
\cite{Chang:1991bp,Scott:1978nz,Braaten:1994bz}.
The perturbative QCD scheme was first applied  by Bjorken \cite{Bjorken:1977md},
Suzuki \cite{Suzuki:1977km},  Amiri and Ji \cite{Amiri:1986zv}, where 
the best elaborate model is proposed by Suzuki \cite{Suzuki:1985up} which is  
based on the convenient Feynman diagrams and the wave function of the respective heavy meson.
In this model,  the heavy FFs are calculated using a diagram similar to that in Fig.~\ref{ff},
so the analytical expression of FFs depends on the transverse momentum $k_T$ of the initial parton which 
appears as a phenomenological parameter (e.g. see Eq.~(\ref{last1})).
In Ref.~\cite{Nejad:2013vsa}, using this model we presented an exact
analytical expression of the initial scale FF for c-quark to split into S-wave $D^+/D^0$ mesons to LO.
To check our results, we showed that there is a good consistency between our result and the current well-known phenomenological
models and  also with the experimental data form BELLE and CLEO.
In Ref.~\cite{Sepahvand:2010zz}, using this model authors have calculated the FFs for a heavy quark Q to fragment
into the vector $Q\bar{Q}(^{3}S_1)$ and pseudoscalar $Q\bar{Q}(^{1}S_0)$ heavy quarkonia to LO. 
Using the same model, authors in Ref.~\cite{GomshiNobary:2003sf} have computed the fragmentation of a heavy quark Q into a QQQ-system
with three identical flavor, i.e. $D_{b\rightarrow \Omega_{bbb}}(z, \mu_0)$ and $D_{c\rightarrow \Omega_{ccc}}(z, \mu_0)$.
Here, $D$ stands for the fragmentation function and $z$ is the fragmentation 
parameter which refers  to the energy fraction of the initial parton which is taken away by the detected hadron and $\mu_0$ is
the initial factorization scale.

In high energy processes, the large contribution of quarkonium production results from gluon fragmentation \cite{Roy:1994ie,Falk:1993rj}.
This is confirmed by the comparison between the theoretical predictions and the experimental
measurements of the heavy quarkonium cross sections.
For this reason, using the Suzuki's model
 we focus on the gluon fragmentation and drive  an analytical
form of the FFs for a gluon to fragment
into the vector $Q\bar{Q}(^{3}S_1)$ and pseudoscalar $Q\bar{Q}(^{1}S_0)$ heavy quarkonia.
To impose the spin effects of the heavy quarkonia into the FFs
we apply three different scenarios.
Comparison of all scenarios shows that the fragmentation probability of 
the gluon into the spin-triplet quarkonium is the biggest one. 

This paper is organized as follows.
In Sec.~\ref{sec:one}, we explain our theoretical model to calculate the FFs in detail.
We discuss the use of pQCD in calculating  
the fragmentation of a charm quark into the heavy charmonium $H_c$.
In Sec.~\ref{sec:two} the analytical expressions for the polarized and unpolarized FFs in three different scenarios 
are presented. Next we shall present our numerical results for the gluon FFs.
Our conclusion is summarized in Sec.~\ref{sec:three}.

\section{Theoretical determination of FFs: Perturbative QCD model}
\label{sec:one}

\begin{figure}
\begin{center}
\includegraphics[width=0.6\linewidth,bb=199 580 392 700]{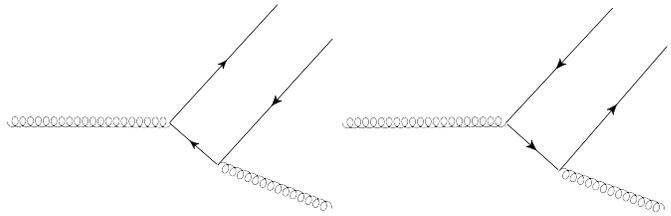}
\caption{\label{graphs}%
Feynman diagrams for the process $g\rightarrow H_c(c\bar c)$ at LO.}
\end{center}
\end{figure}

\begin{figure}
\begin{center}
\includegraphics[width=0.42\linewidth,bb=199 440 392 729]{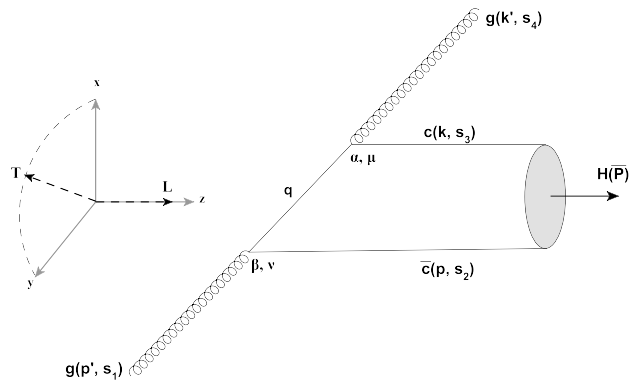}
\caption{\label{ff}%
Formation of a heavy charmonium. A gluon forms a bound state $c\bar c$
with a gluon produced through a single c-quark.}
\end{center}
\end{figure}

As  is pointed out in Refs.~\cite{Braaten:1993rw,Chang:1991bp}
the fragmentation function for hadrons containing a heavy quark $Q$
or a heavy antiquark $\bar Q$ can be computed theoretically using perturbative QCD (pQCD).
The first theoretical effort to illustrate the hadroproduction procedure by a heavy quark was made by Bjorken \cite{Bjorken:1977md}
by using a naive quark-parton model. He found out that the inclusive distribution of  heavy hadron  should peak
nearly at $z=1$, where $z$ refers to  the scaled energy variable. This property
is important for heavy quarks for which the peak of heavy quark FF happens closer to $z=1$.
The pQCD model was followed by Suzuki \cite{Suzuki:1977km},  Amiri and Ji \cite{Amiri:1986zv}.\\
Here, using the Suzuki's model we focus on the gluon fragmentation function and  obtain an analytical form of the gluon 
fragmentation into a spin-singlet and -triplet S-wave heavy quarkonium by employing
three different scenarios. Our results shall be compared with the unpolarized one.
To this end, we consider  a special example: $g\rightarrow H_c(=c\bar{c})$ for which 
the respective Feynman diagrams shown in Fig.~\ref{graphs}.
First, we consider Fig.~\ref{ff} where the basic process of gluon fragmentation into the charmonium bound state
along with the spins and the four-momenta of meson and partons are shown.
Following Ref.~\cite{Suzuki:1985up}, we  adopt the infinite momentum frame where
the fragmentation  parameter in the usual form $z=(E^H+p_{||}^H)/(E^g+p_{||}^g)$ is 
reduced to the more popular one as 
\begin{eqnarray}\label{parameter}
z=\frac{E^H}{E^g}\cdot
\end{eqnarray}
With the large heavy quark mass, the relative motion
of the heavy quark pair inside the quarkonium is effectively nonrelativistic \cite{Ma:2013yla},
so the squared relative velocity of the heavy quark pair in the quarkonium rest frame is 
$v^2\approx 0.22$ for the $J/\psi$ and $v^2\approx 0.1$ for the $\Upsilon$ \cite{Bodwin:2012xc}.
Here, according to the Lepage-Brodsky's approach \cite{Lepage:1980fj}, by neglecting the relative motion of  the heavy
quark pair inside the quarkonium we assume, for simplicity, that
$Q$ and $\bar Q$ are emitted collinearly with each other and move along the $Z$-axes.
Indeed, the Fermi motion of the constituent quarks in the bound state is neglected.

In the Suzuki's model, the FF for the production of S-wave bound state is defined as \cite{Suzuki:1985up,GomshiNobary:1994eq}
\begin{eqnarray}\label{first}
D_g^{H_c}(z, \mu)=\int d^3\bold{\bar P}d^3\bold{k^\prime} |T_M|^2\delta^3(\bold{k^\prime}+\bold{\bar P}-
\bold{p^\prime}),
\end{eqnarray}
where $T_M$ is the probability amplitude of the meson production which 
is expressed as the convolution of the hard scattering amplitude $T_H$
and the process-independent distribution amplitude $\Phi_M$.
The hard amplitude $T_H$ can be computed perturbatively from partonic
subprocesses, and the distribution amplitude $\Phi_M$  contains 
the bound state nonperturbative dynamic of outgoing meson.
This convolution is expressed as
\begin{eqnarray}\label{base}
T_M=\int [dx_i] T_H(x_i,Q^2) \Phi_M(x_i, Q^2),
\end{eqnarray}
where, $[dx_i]=dx_1dx_2\delta(1-x_1-x_2)$ and $x_i$'s are the momentum fractions
carried by the constituent quarks.
This scheme,  introduced in \cite{Adamov:1997yk,brodsky}, is applied to absorb the soft behavior of the bound state into the  hard scattering amplitude $T_H$.
The amplitude $T_H$ is, in essence, the partonic cross section to produce a heavy quark-antiquark ($Q\bar Q$)
pair with certain quantum numbers that, in the old fashioned perturbation theory, is written as
\begin{eqnarray}\label{first2}
T_H=\frac{4\pi\alpha_s m_c^2 C_F}{2\sqrt{2\bar{P}_0 k_0^\prime{p}_0^\prime}}\frac{\Gamma}{(\bar{P}_0+k_0^\prime-p_0^\prime)},
\end{eqnarray}
where $C_F$ is the color factor, $\alpha_s$ is  the strong coupling constant
and $\Gamma$ represents an appropriate combination of the quark propagator and spinorial 
parts of the amplitude as
\begin{eqnarray}\label{second}
\Gamma=G\bigg\{\bar{u}(k, s_3) \displaystyle{\not}\epsilon_4^\star 
 (\displaystyle{\not}{q}+m_c)\displaystyle{\not}\epsilon_1 v(p, s_2)\bigg\}.
\end{eqnarray}
Here, $\epsilon$ is the polarization vector of gluon and $G=1/(q^2-m_c^2)=1/(2k.k^\prime)$
is proportional to the quark propagator. 
In (\ref{base}),  $\Phi_M$ is the process-independent probability amplitude to find the quarks which are collinear in
the mesonic bound state and  contains the  nonperturbative dynamic of the
mesonic bound state.
In general, the distribution amplitude is related to the mesonic  wave function $\Psi_M$  by
\begin{eqnarray}\label{formul}
\Phi_M(x_i, Q^2)=\int [d^2 \bold{q_{\bot i}}]\Psi_M(x_i, \bold{q_{\bot i}})\Theta( \bold{q_{\bot i}}^2<Q^2),
\end{eqnarray}
where
\begin{eqnarray}
 [d^2 \bold{q_{\bot i}}]=2 (2\pi)^3\delta\big[\sum_{j=1}^2  \bold{q_{\bot j}}\big]\prod_{i=1}^2\frac{d^2 \bold{q_{\bot i}}}{2(2\pi)^3}.
\end{eqnarray}
In the equation above, $ \bold{q_{\bot i}}$ refers to the  transverse momentum of constituent quarks.  
A hadronic wave function $\Psi_M$ which is the  nonrelativistic limit of the Bethe-Salpeter equation with the QCD kernel is given in \cite{brodsky}, so
that by working in the infinite-momentum frame it can be estimated as a delta function \cite{Nejad:2013vsa}.
Therefore, the distribution amplitude  for a S-wave heavy meson at large $Q^2$, reads \cite{Amiri:1985mm}
\begin{eqnarray}\label{wave}
\Phi_M\approx\frac{f_M}{2\sqrt{3}} \delta(x_1-\frac{m_c}{M}),
\end{eqnarray}
where $M$ is the heavy meson mass and $f_M=\sqrt{12/M}|\Psi(0)|$ is the meson decay constant which is related to the the nonrelativistic S-wave function $\Psi(0)$
at the origin.
The delta-function form is convenient for our assumption where the constituent quarks are emitted collinearly, without
any transverse momentum.
Substituting Eqs.~(\ref{first2}) and (\ref{wave}) in (\ref{base}) and carrying out the necessary
integrations, the probability amplitude (\ref{base}) is found as 
\begin{eqnarray}\label{third}
T_M=\frac{\pi \alpha_s m_c^2 f_M C_F}{\sqrt{6\bar{P}_0 k_0^\prime{p}_0^\prime}}\frac{\Gamma}{(\bar{P}_0+k_0^\prime-p_0^\prime)}.
\end{eqnarray}
Now the fragmentation function (\ref{first}) reads
\begin{eqnarray}\label{int}
D_g^{H_c}(z, \mu_0)&=&\frac{1}{6}[\pi\alpha_s f_M C_F m_c^2]^2\nonumber\\
&&\hspace{-1cm}\times\int\frac{\bar{\Gamma}\Gamma\delta^3(\bold{k}^\prime+\bold{\bar{P}}-\bold{p}^\prime)}{(\bar{P}_0 p_0^\prime k_0^\prime)
(\bar{P}_0+k_0^\prime-p_0^\prime)^2}
d^3\bold{\bar{P}} d^3\bold{k}^\prime.\nonumber\\
\end{eqnarray}

\subsection{kinematics}

To proceed our discussion, considering Fig.~\ref{ff} we need to specify our kinematics. We choose an infinite
momentum frame in which the relevant four-momentum of the initial gluon is set as $p_\mu^\prime =[p_0^\prime; \bold{k_T}, p_L^\prime]$.
The  transverse momentum of the initial gluon is only carried by the final state gluon which produces a jet and
its four-momentum is written as $k_\mu^\prime =[k_0^\prime; \bold{k_T}, k_L^\prime]$. We
let the produced heavy quarkonium  moves in the fragmentation axis ($Z$-axis), which is defined as the direction of the three-momentum
of the heavy quark Q in the laboratory frame after production.
The four-momentum of quarkonium is set as $\bar P_\mu=[\bar P_0; \bold{0}, \bar P_L]$.
The constituents of the quarkonium after creation move along the fragmentation axis
and their  momenta are defined as $p_\mu=[p_0; \bold{0}, p_L]$ and $k_\mu=[k_0; \bold{0}, k_L]$. 
We also assume that there is only one jet in the final state, since the very high momentum of the
initial parton will predominantly be carried in the forward direction.\\
According to the definition of the fragmentation parameter $z=E^H/E^g=\bar P_0/p_0^\prime$ (\ref{parameter}), the heavy quarkonium takes a fraction $z$ of the
initial gluon energy $p^\prime_0$ and the outgoing gluon takes the
remaining $(1 - z)$. Therefore, the parton energies can be parametrized in terms of the  initial gluon energy;
$p_0=x_1 z p_0^\prime, k_0=x_2 z p_0^\prime,  k_0^\prime=(1-z) p_0^\prime$, 
where $x_1=p_0/\bar{P}_0$ and $x_2=k_0/\bar{P}_0$ are  the quarkonium energy fractions carried by the constituent quarks.
Following Ref.~\cite{GomshiNobary:1994eq,Kolodziej:1994uu}, we also assume  that the contribution of each constituent quark from the meson energy is proportional
to its mass, i.e. $x_1=m_c/M$ and $x_2=m_{\bar c}/M$, where $M$ is the heavy quarkonium  mass. Note that for charmonium and bottomonium states we have
$m_1=m_2=m_Q$ and $M\cong 2m_Q$, then one has $x_1=x_2\cong 1/2$.\\

\subsection{Unpolarized fragmentation functions}
 
To obtain an  analytical form of FF for gluon to split into the unpolarized charmonium state, in Eq.~(\ref{int}) 
one has to perform  an average over the spin of the initial gluon and a sum over the colors and spins. Thus the mean amplitude squared reads
\begin{eqnarray}
\overline{\Gamma^2}=\overline{|\Gamma \bar{\Gamma}|}=\frac{1}{1+2s_g}\sum_{s, c}\Gamma\bar{\Gamma},
\end{eqnarray}
where $s_g$ is the spin of the initial gluon. Then we obtain
\begin{eqnarray}\label{tensor}
\sum_{s,c}\Gamma\bar{\Gamma}&=&G^2Tr[(\displaystyle{\not}{k}+m_c)\gamma^\mu(\displaystyle{\not}{k}+\displaystyle{\not}{k^\prime}+m_c)
\gamma^\nu(\displaystyle{\not}{p}-m_c)\nonumber\\
&&\times \gamma_\nu (\displaystyle{\not}{k}+\displaystyle{\not}{k^\prime}+m_c)\gamma_\mu]
=-32G^2\bigg\{2m_c^4+\nonumber\\
&&m_c^2[2k.k^\prime+k.p+p.k^\prime]-(p.k^\prime)(k.k^\prime)\bigg\},
\end{eqnarray}
where, we applied the energy projection operators
\begin{eqnarray}\label{operr}
\Lambda^+(k)&=&\sum_{S_3} u(k, s_3) \bar{u}(k, s_3)=(\displaystyle{\not}{k}+m_c),\nonumber\\
\Lambda^-(p)&=&\sum_{S_2} v(p, s_2) \bar{v}(p, s_2)=(\displaystyle{\not}{p}-m_c).
\end{eqnarray}

\subsection{Polarized fragmentation functions: First scenario}

Considering $v(p)$ and $\bar{u}(k)$ as the Dirac spinors of the quarks forming the 
charmonium bound states,  in the nonrelativistic approximation the
projection operator can be defined as \cite{Kuhn:1979bb,Kolodziej:1994uu}
\begin{eqnarray}
\Lambda_{S,S_z}(\bar{P})=v(p)\bar{u}(k)=\frac{f_M}{\sqrt{48}}(\displaystyle{\not}{\bar{P}}-M)\Pi_{SS_z},
\end{eqnarray} 
where $M$ and $\bar{P}$ are the mass and the four-momentum of  meson bound state, $f_M$ is the meson decay constant
and $\Pi_{S S_z}$ is the appropriate spin projection operator; $\Pi_{00}=\gamma_5$
for pseudoscalar and $\Pi_{1S_z}=\displaystyle{\not}{\epsilon}$ for vector states.
The spin content of meson is then given by either $\gamma_5$ or $\displaystyle{\not}{\epsilon}(S_z)$, as could well be expected.
Therefore, the spinorial part of the amplitude (\ref{second}) for formation of the pseudoscalar 
($S=0$) and the vector ($S= 1$) charmonium states may be
presented in the following forms
\begin{eqnarray}\label{firstscenarion}
\Gamma^P&\propto&G\bigg\{(\displaystyle{\not}{\bar{P}}-M)\gamma_5 \displaystyle{\not}\epsilon_4^\star 
(\displaystyle{\not}{q}+m_c)\displaystyle{\not}\epsilon_1\bigg\},\nonumber\\
\Gamma^V&\propto&G\bigg\{(\displaystyle{\not}{\bar{P}}-M)\displaystyle{\not}\epsilon\displaystyle{\not}\epsilon_4^\star 
(\displaystyle{\not}{q}+m_c)\displaystyle{\not}\epsilon_1\bigg\},
\end{eqnarray}
where $q=k+k^\prime$ is the energy-momentum of the virtual intermediate quark, $\epsilon$ is the polarization four-vector of meson which may be in a longitudinal state
$\epsilon^{(L)\mu}=\epsilon^\mu(\bar{P}, \lambda=0)$ or a transverse state $\epsilon^{(T)\mu}=\epsilon^\mu(\bar{P}, \lambda=\pm 1)$.
These components satisfy the following relations
\begin{eqnarray}
\epsilon(\bar{P}, \lambda).\bar{P}=0\quad &,& \quad \bold\epsilon(\bar{P}, \lambda).\bold\epsilon^\star(\bar{P}, \lambda^\prime)=-\delta_{\lambda, \lambda^\prime}\nonumber\\
\bold\epsilon^{(T)}.\bold{\bar{P}}=&0&=\bold\epsilon^{(L)}\times\bold{\bar{P}}.
\end{eqnarray}
Therefore, for a quarkonium for which $\bar P^\mu=[\bar P_0; \bold{0}, \bar P_L]$, the polarization four-vector (or the spin wave function)
is expressed as 
\begin{eqnarray}
\epsilon^{(L)\mu}&=&\frac{1}{M}(\bar{P}_L; 0, 0, \bar{P}_0)=\frac{1}{M}(k_L+p_L; 0, 0, p_0+k_0),\nonumber\\
\epsilon^{(T)\mu}&=&\frac{1}{\sqrt{2}}(0; \mp 1, -i, 0).
\end{eqnarray}
Now to obtain an  analytical form of FF for  the polarized charmonium, in (\ref{int})
we perform  an average over the spin of the initial gluon and a sum over the colors and the spins of  gluons.
The result reads
\begin{eqnarray}\label{sq}
|\Gamma^{V,P}|^2=\frac{1}{1+2s_g}\sum_{s_g, c_g}\Gamma^{V,P}\bar{\Gamma}^{V,P},
\end{eqnarray}
then we find
\begin{eqnarray}
\sum_{s_g}\Gamma^{P}\bar{\Gamma}^{P}&=&128G^2\bigg\{2m_c^4-Mm_c^3+2(k.p)(k.k^\prime)\nonumber\\
&&+M^2(k.k^\prime)+m_c^2\big[M^2+2k.k^\prime+2k.p\big]\nonumber\\
&&-Mm_c\big[k.k^\prime+k.p+p.k^\prime\big]\bigg\},\nonumber\\
\sum_{s_g}\Gamma^{V}\bar{\Gamma}^{V}&=&128G^2\bigg\{-m_c^2M^2+2m_c^2(p^\prime.k^\prime)-M^2(k.k^\prime)\nonumber\\
&&+2(k.k^\prime)(k^\prime.p^\prime)-Mm_c\big[k.k^\prime-k.p^\prime-p^\prime.k^\prime\nonumber\\
&&+2(\epsilon.k^\prime)^2+2(\epsilon.k)(\epsilon.k^\prime)-2(\epsilon.k)(\epsilon.p^\prime)-\nonumber\\
&&2(\epsilon.p^\prime)(\epsilon.k^\prime)\big]\bigg\}.
\end{eqnarray}

\subsection{Polarized fragmentation functions: Second scenario}

Following Ref.~\cite{Guberina:1980dc}, in the nonrelativistic approximation the spin
projection operator can be also defined as 
\begin{eqnarray}
\Lambda_{S,S_z}(p,k)&=&\nonumber\\
&&\hspace{-1cm}\sqrt{\frac{3}{m}}\sum_{S_2, S_3}\left\langle \frac{1}{2}S_2; \frac{1}{2}S_3|S, S_z\right\rangle v(p+q)\bar{u}(k-q),\nonumber\\
\end{eqnarray} 
where $\left\langle S_2/2;S_3/2|S, S_z\right\rangle$ 
are the Clebsch-Gordan coefficients and $S_2$, $S_3$ and $q$ are the spins and the relative four-momentum  of the constituent quarks, respectively.
These matrices can be written in a simple covariant form up
to terms which are of $q^2$-order in the quarkonium rest frame. Such terms are
irrelevant in the nonrelativistic approximation which we are
adopting. Therefore one finds \cite{Guberina:1980dc}, 
\begin{eqnarray}\label{operator}
\Lambda_{S,S_z}(p,k)=\sqrt{\frac{3}{8m^2}}[(\displaystyle{\not}{k}-\displaystyle{\not}{q}+m_2)\Pi_{S, S_z}(-\displaystyle{\not}{p}-\displaystyle{\not}{q}+m_1)],\nonumber\\
\end{eqnarray} 
where $\Pi_{S, S_z}$ is as before. Ignoring the Fermi motion, 
so that the constituent quarks  will fly 
together in  parallel, Eq.~(\ref{operator}) can be simplified as
\begin{eqnarray}
\Lambda_{S,S_z}(p,k)\propto(\displaystyle{\not}{k}+m_2)\Pi_{S, S_z}(\displaystyle{\not}{p}-m_1).
\end{eqnarray} 
In our calculation $m_1=m_2=m_c$. 
Thus, the amplitude squared $\Gamma\bar \Gamma$ (\ref{sq}) in the second scenario is given by
\begin{eqnarray}
\sum_{s_g}\Gamma^{V,P}\bar{\Gamma}^{V,P}&=&G^2Tr\bigg[(\displaystyle{\not}{k}+m_c)\Pi_{S, S_z}(\displaystyle{\not}{p}-m_c)\gamma^\mu(\displaystyle{\not}{q}+m_c)\nonumber\\
&&\hspace{-1.5cm}\times\gamma^\nu\gamma_\nu(\displaystyle{\not}{q}+m_c)\gamma_\mu(\displaystyle{\not}{p}-m_c)\Pi_{S, S_z}(\displaystyle{\not}{k}+m_c)\bigg],
\end{eqnarray}
where $\Pi_{S, S_z}=\gamma_5$ for $c\bar{c}(^{1}S_0)$ and $\Pi_{S, S_z}=\displaystyle{\not}{\epsilon}$ for $c\bar{c}(^{3}S_1)$.
Using the traditional trace technique, the necessary traces are
evaluated and the results are expressed as the dot products of four-vectors.

\subsection{Polarized fragmentation functions: Third scenario}

In the third scenario the spin projection operator is defined as 
\begin{eqnarray}
\Lambda_{S,S_z}(p,k)\propto (\displaystyle{\not}{p}+m_c)\Pi_{S, S_z}.
\end{eqnarray} 
This scenario is defined in \cite{Guberina:1980dc} and applied in \cite{Sepahvand:2010zz} to obtain the spin dependent fragmentation functions for a heavy quark $Q$ to fragment
into vector $Q\bar{Q}(^{3}S_1)$ and pseudoscalar $Q\bar{Q}(^{1}S_0)$ heavy quarkonium to leading order perturbative
QCD. Therefore, 
\begin{eqnarray}
\sum_{s_g}\Gamma^{V,P}\bar{\Gamma}^{V,P}&=&G^2Tr\bigg[(\displaystyle{\not}{p}+m_c)\Pi_{S, S_z}\gamma^\mu(\displaystyle{\not}{q}+m_c)\gamma^\nu\gamma_\nu\nonumber\\
&&\times(\displaystyle{\not}{q}+m_c)\gamma_\mu\Pi_{S, S_z}(\displaystyle{\not}{p}+m_c)\bigg].
\end{eqnarray}
Now, the spinorial part of the amplitude to produce the pseudoscalar 
and vector  charmonium states read as 
\begin{eqnarray}
\sum_{s_g}\Gamma^{P}\bar{\Gamma}^{P}&=&128G^2m_c^2\Big\{2m_c^2+2(k.k^\prime)+k.p+p.k^\prime\Big\},\nonumber\\
\sum_{s_g}\Gamma^{V}\bar{\Gamma}^{V}&=&-128G^2\bigg\{2(k.k^\prime)\bigg[m_c^2-p^\prime.k^\prime-k.p^\prime+k.k^\prime\bigg]\nonumber\\
&&+m_c^2\bigg[m_c^2-2(k.\epsilon)^2-2(k^\prime.\epsilon)^2-k.p^\prime-k^\prime.p^\prime\nonumber\\
&&+2(k^\prime.\epsilon)(p^\prime.\epsilon)+(k.\epsilon)\bigg(2p^\prime.\epsilon-4k^\prime.\epsilon\bigg)\bigg]\bigg\}.\nonumber\\
\end{eqnarray}

\section{Analytical results}
\label{sec:two}

To obtain the FFs for unpolarized and polarized quarkonia,
using the kinematics, first we put the dot products of the relevant four-vectors in the following forms:
\begin{eqnarray}
2k.p&=&\frac{M^2}{8},\nonumber\\
2p^\prime.k^\prime &=&\frac{z^2}{1-z}k_T^2,\nonumber\\
2\epsilon_L.k &=&2\epsilon_L.p=0,\nonumber\\
2\epsilon_L.p^\prime &=&-\frac{M}{z}+\frac{zk_T^2}{M},\nonumber\\
2\epsilon_T.p^\prime&=&-\sqrt{2}k_T(-1+i),\nonumber\\
2\epsilon_T.k^\prime&=&-\sqrt{2}k_T(-1-i),\nonumber\\
2k.p^\prime &=&2p.p^\prime=\frac{zm}{M}k_T^2+\frac{mM}{z},\nonumber\\
2\epsilon_L.k^\prime &=&-\frac{M(1-z)}{z}+\frac{zk_T^2}{M(1-z)},\nonumber\\
2p.k^\prime &=&2k.k^\prime=\frac{zm}{(1-z)M}k_T^2+\frac{mM(1-z)}{z}.
\end{eqnarray}
To perform the phase space integrations (\ref{int}), first we consider the following integral
\begin{eqnarray}
\int \frac{d^3\bold{\bar{P}}\delta^3(\bold{\bar{P}}+
\bold{k^\prime}-\bold{p^\prime})}{\bar{P}_0 (\bar{P}_0+k_0^\prime-p_0^\prime)^2}
=\frac{\bar{P}_0}{(M^2+2p^\prime.k^\prime)^2},
\end{eqnarray}
and instead of performing the transverse momentum integration we replace the integration variable
by its average value $ \left\langle k_T^2 \right\rangle$, 
which is a free parameter and can be specified experimentally. Therefore we can write
\begin{eqnarray}
\int{ F(z, k_T^2) d^3 k^\prime}&=&\int{F(z, k_T^2) dk_L^\prime d^2k_T^\prime} \nonumber\\
&\approx & m_c^2 k_0^\prime F(z, \left\langle k_T^{2}\right\rangle).
\end{eqnarray}
In conclusion, considering the contributions of both Feynman diagrams in Fig.~\ref{graphs}, we obtain the
\textit{unpolarized} FF for $g\rightarrow H_c(c\bar{c})$ at the initial scale $\mu_0=2m_c$ as
\begin{eqnarray}\label{last1}
D_{g\rightarrow H_c}(z, \mu_0)&=&\frac{N z}{g(z, \left\langle k_T^{2}\right\rangle)}\times\nonumber\\
&&\hspace{-2.5cm}\bigg\{\bigg[\frac{z^2k_T^{2}+M^2(1-z)^2}{Mz(1-z)}-3m_c\bigg]^2-21m_c^2\bigg\},
\end{eqnarray}
where,
\begin{eqnarray}
g(z, \left\langle k_T^{2}\right\rangle)=\bigg[M^2+\frac{z^2k_T^2}{1-z}\bigg]^2
\bigg[\frac{z^2k_T^{2}+M^2(1-z)^2}{Mz(1-z)}\bigg]^2,\nonumber\\
\end{eqnarray}
and $N$ is proportional to $(\pi C_F m_c^3\alpha_s f_M)^2$ but it is related to the normalization condition \cite{Amiri:1986zv,Suzuki:1985up}.

The \textit{polarized} FFs, using the \textit{first scenario} are written as
\begin{eqnarray}\label{last2}
D_{g}^{P}(z, \mu_0)&=&\frac{N z}{g(z, \left\langle k_T^{2}\right\rangle)}
(64m_c)\bigg[3m_c^2+(M-m_c)^2\bigg]\nonumber\\
&&\times\bigg[2m_c+\frac{zk_T^2}{M(1-z)}+\frac{M(1-z)}{z}\bigg],\nonumber\\
D_{g}^{T}(z, \mu_0)&=&\frac{N z}{g(z, \left\langle k_T^{2}\right\rangle)} (64m_c)\bigg[\frac{z^3k_T^4}{M(1-z)^2}-m_cM^2\nonumber\\
&&-M^3\frac{1-z}{z}+k_T^2(4M+\frac{m_c}{1-z}z^2)\bigg],\nonumber\\
D_{g}^{L}(z, \mu_0)&=&\nonumber\\
&&\hspace{-1cm}\frac{N z}{g(z, \left\langle k_L^{2}\right\rangle)}\frac{64m_c}{z}\bigg[\frac{z^2k_T^2}{(1-z)}(m_c-M)-m_cM^2\bigg].\nonumber\\
\end{eqnarray}
Note that, the fragmentation function for a vector charmonium $H_c(^{3}S_1)$ is the sum of the longitudinal and twice the
transverse components, i.e.,
\begin{eqnarray}\label{sum}
D_{g}^{V}(z, \mu_0)=2D_{g}^{T}+D_{g}^{L},
\end{eqnarray}
where $T$ and $L$ refer to the longitudinal and the transverse polarization of vector charmonium.

The polarized FFs, using the \textit{second scenario} may be written as
\begin{eqnarray}\label{last3}
D_{g}^{P}(z, \mu_0)&=&\frac{N z}{g(z, \left\langle k_T^{2}\right\rangle)}(768m_c^5)\nonumber\\
&&\times\bigg[\frac{zk_T^2}{M(1-z)}+2m_c-\frac{M(z-1)}{z}\bigg],\nonumber\\
D_{g}^{T}(z, \mu_0)&=&\frac{N z}{g(z, \left\langle k_T^{2}\right\rangle)} \frac{64m_c}{z}\bigg[\frac{z^6k_T^6(m_c-M)^2}{M^3(z-1)^3}\nonumber\\
&&\hspace{-1.8cm}+\frac{z^2k_T^4}{M^2(1-z)^2}\bigg\{-6z^2m_c^3+Mm_c^2(z^3+11z^2-12z\nonumber\\
&&\hspace{-1.8cm}+8)-2m_cM^2z(z^2+2z-2)+z^2M^3(z-1)\bigg\}+\nonumber\\
&&\hspace{-1.8cm}\frac{m_ck_T^2}{M(z-1)}\bigg\{-2M^3z^2(z-1)+4z^2m_c^3(1+2z)-\nonumber\\
&&\hspace{-1.8cm}2Mm_c^2z(7z^2-8z+8)+m_cM^2(8z^3-11z^2\nonumber\\
&&\hspace{-1.8cm}+12z-8)\bigg\}-m_c^2\bigg\{8zm_c^3+4Mm_c^2(1-3z)+\nonumber\\
&&\hspace{-1.8cm}6m_cM^2(z-1)-M^3(z-1)\bigg\}\bigg],\nonumber\\
D_{g}^{L}(z, \mu_0)&=&\frac{N z}{g(z, \left\langle k_L^{2}\right\rangle)}\frac{64m_c}{z}\bigg[\frac{z^6k_T^6}{M^3(z-1)^3}\bigg\{M^2+\nonumber\\
&&\hspace{-1.8cm}(z-3)m_c(M-m_c)\bigg\}-\frac{z^4k_T^4}{M^2(1-z)^2}\bigg\{2m_c^3(z+3)\nonumber\\
&&\hspace{-1.8cm}-Mm_c^2(2z+5)+3zm_cM^2+M^3(1-z)\bigg\}+\nonumber\\
&&\hspace{-1.8cm}\frac{m_cz^2k_T^2}{M(z-1)}\bigg\{(8z+4)m_c^3-10Mm_c^2z+\nonumber\\
&&\hspace{-1.8cm}M^2m_c(7z-5)+M^3(3-2z)\bigg\}-m_c^2\bigg\{ 8zm_c^3+\nonumber\\
&&\hspace{-1.8cm}4Mm_c^2(1-3z)+2M^2m_c(4z-3)+M^3(3-2z)\bigg\}\bigg].\nonumber\\
\end{eqnarray}

\begin{figure}
\begin{center}
\includegraphics[width=1\linewidth,bb=88 610 322 769]{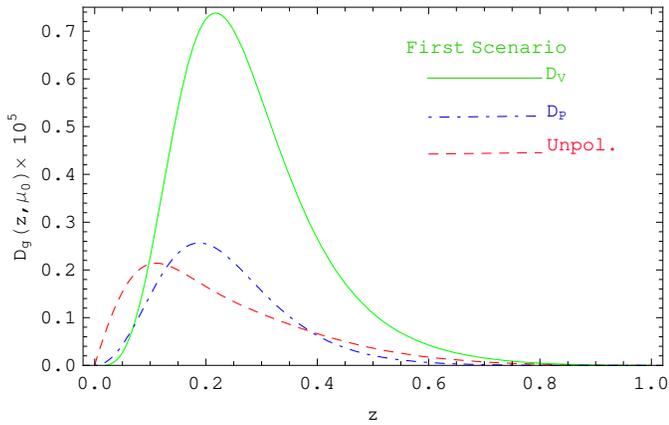}
\caption{\label{fig3}%
The fragmentation functions for the vector (solid line) and the pseudoscalar (dot-dashed line) charmoniums in  
the first scenario.
For comparison the unpolarized one (dashed line) is also shown. The initial scale is $\mu_0=2m_c$}
\end{center}
\end{figure}

\begin{figure}
\begin{center}
\includegraphics[width=1\linewidth,bb=88 610 322 769]{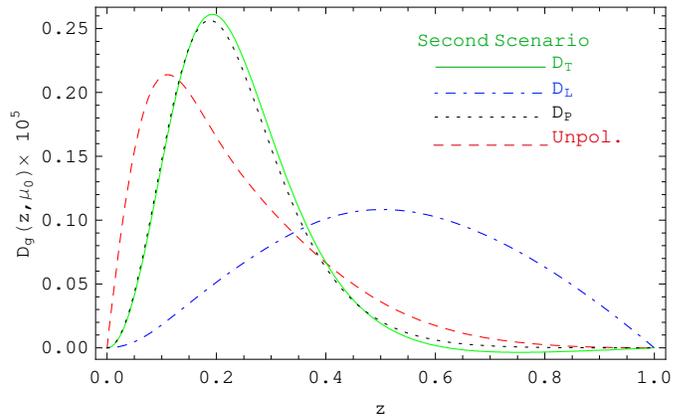}
\caption{\label{fig4}%
The longitudinal (dot-dashed line) and transverse (solid line) components of the vector charmonium fragmentation function 
in the second scenario. The pseudoscalar (dots) and the unpolarized (dashed line) FFs are also shown.}
\end{center}
\end{figure}

\begin{figure}
\begin{center}
\includegraphics[width=1\linewidth,bb=88 610 322 769]{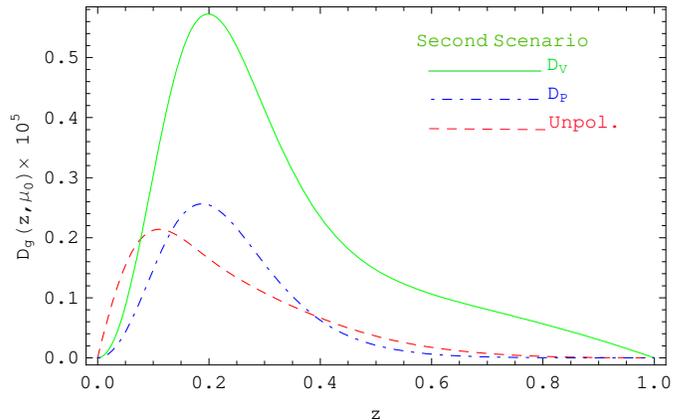}
\caption{\label{fig5}%
As in Fig.~\ref{fig3} but in the second scenario.}
\end{center}
\end{figure}

\begin{figure}
\begin{center}
\includegraphics[width=1\linewidth,bb=88 610 322 769]{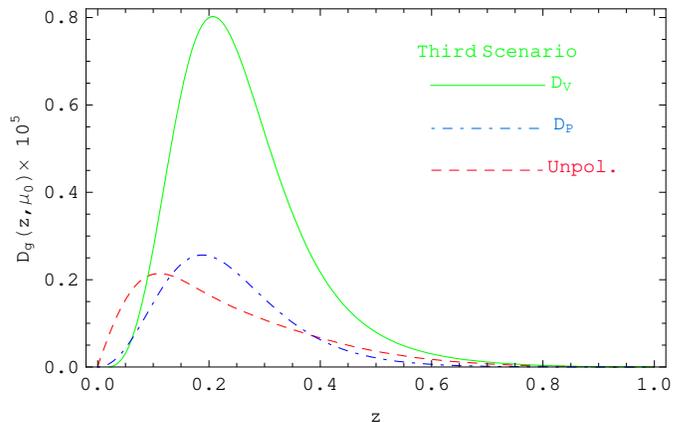}
\caption{\label{fig6}%
As in Fig.~\ref{fig3} but in the third scenario.}
\end{center}
\end{figure}

And the polarized FFs in the \textit{third scenario} are expressed as
\begin{eqnarray}\label{last40}
D_{g}^{P}(z, \mu_0)&=&\frac{N z}{g(z, \left\langle k_T^{2}\right\rangle)}(192 m_c^3)\nonumber\\
&&\times\bigg[\frac{zk_T^2}{M(1-z)}+2m_c-\frac{M(z-1)}{z}\bigg],\nonumber\\
D_{g}^{T}(z, \mu_0)&=&\frac{N z}{g(z, \left\langle k_T^{2}\right\rangle)} \frac{64m_c}{z}\bigg[\frac{z^4k_T^4(M-m_c)}{M^2(z-1)^2}\nonumber\\
&&+\frac{zk_T^2}{M(z-1)}\bigg\{-2Mm_c((1+z)^2-3)\nonumber\\
&&+M^2z(z-1)+m_c^2z(1+z)\bigg\}-m_c\bigg\{\nonumber\\
&&2zm_c^2+M^2(z-1)+Mm_c(1-2z)\bigg\}\bigg],\nonumber\\
\end{eqnarray}
and
\begin{eqnarray}\label{last41}
D_{g}^{L}(z, \mu_0)&=&\frac{N z}{g(z, \left\langle k_L^{2}\right\rangle)}\frac{64m_c}{z}\bigg[\frac{z^4k_T^4}{M(z-1)^2}+\nonumber\\
&&\hspace{-0.5cm}\frac{z^2k_T^2}{M(z-1)}\bigg\{m_c^2(1+z)-3m_cMz+\nonumber\\
&&\hspace{-0.5cm}M^2(z-1)\bigg\}-m_c^2\bigg(M+2z(m_c-M)\bigg)\bigg].\nonumber\\
\end{eqnarray}
We are now in a position to present our phenomenological predictions for the gluon fragmentation into the spin-singlet
and -triplet S-wave charmoniums, by performing a numerical analysis.
In general, fragmentation function $D_{g\rightarrow H_c}(z,\mu^2)$ depends on both the fragmentation parameter
$z$ and factorization scale $\mu$.
The scale $\mu$  is generally arbitrary, but in  a high energy process where a jet is produced with transverse momentum $ k_T$,
large logarithms of $k_T/\mu$ in the partonic cross section (perturbative part of the hadroproduction)
can be avoided by choosing  $\mu$  on the order of $ k_T$.
The functions (\ref{last1}, \ref{last2}, \ref{last3}-\ref{last41}) should be regarded as models for the gluon fragmentation 
 into the polarized and unpolarized charmoniums at a scale $\mu$ of order $m_Q$.
For values of $\mu$ much larger than $\mu_0$, the initial FFs should be evolved from the
scale $\mu_0$ to the scale $\mu$ using the Altarelli-Parisi equations \cite{dglap1,dglap2, dglap3}.\\
For numerical results, we take $m_c=1.5$ GeV,  $\alpha_s(2m_c)=0.26$ and $f_M(c\bar{c})=0.48$ GeV \cite{Gomsh}.
The initial scale of fragmentation
is set to $\mu_0=2m_c$ and the transverse momentum of the initial gluon is selected as $\left\langle k_T\right\rangle=10$ GeV. 
In Fig.~\ref{fig3}, using the first scenario our predictions for the polarized charmonium FFs, i.e. $g\rightarrow H_c(^{1}S_0,^{3}S_1)$, and
the unpolarized one are shown. Here, $D_V$ is the convenient summation of the longitudinal and transverse fragmentation functions, see Eq.~(\ref{sum}). 
In Fig.~\ref{fig4}, using the second scenario the behavior of the fragmentation functions for the longitudinally and transversely
components of vector  meson are shown. Those are also compared with the  pseudoscalar and the unpolarized ones, 
while for the unpolarized case a summation is going over the spin of the constituent heavy quarks in which case we apply
the energy projection operators defined in (\ref{operr}). 
The fragmentation functions shown in Fig.~\ref{fig4} are also compared in Fig.~\ref{fig5} by taking $D^V=2D^T+D^L$. 
In Fig.~\ref{fig6}, the same comparison is done but using the third scenario.
As is seen, in all scenarios the fragmentation  probability of the gluon  into the spin-triplet
charmonium is the biggest one.
The position of the fragmentation peak is almost the same in all scenarios but the third scenario shows  a bigger peak.\\
Our result shown in Fig.~\ref{fig6}, is in reasonable agreement with the result presented
in Fig.~3 of Ref.~\cite{Qi:2007sf}, when the non-covariant definition
of fragmentation parameter (\ref{parameter}) is applied.
In both results, the peak position of the fragmentation function occurs at $z\approx 0.22$ 
when $p_T=10$ GeV is considered, and
the maximum value of the vector charmonium FF is $D_g^V\approx 0.8\times 10^{-5}$.
Note that in \cite{Qi:2007sf}, authors have applied the Braaten's approach which is completely different
with the scheme used in our work. \\
Besides the $g\rightarrow H_c$ FF itself, also its first moment is of phenomenological interest.
It corresponds to the $g\rightarrow H_c$ branching fraction
\begin{eqnarray}
B(\mu)=\int_0^1 dz D_{g}^{H_c}(z, \mu).
\end{eqnarray}
Using the third scenario, our result for the $g\rightarrow H_c(^{3}S_1)$ branching fraction
is  $B(2m_c)=3.01\times 10^{-6}$ which
can be compared with the result presented in \cite{Braaten:1993rw} where $B(2m_c)=3.2\times 10^{-6}$.\\
By these comparisons, it seems that the third scenario
is more suitable than the other ones. However, 
the advantage of the projection operator in the first scenario is that it defines the 
bound state mass of heavy meson in our calculations.\\
Our results can be directly applied to the S-wave bottomonium sates $\Upsilon (^{3}S_1)$ and $\eta_b(^{1}S_0)$, except 
that $m_c$ is replaced by $m_b=4.5$ GeV and the decay constant $f_M(b\bar{b})=0.33$ GeV \cite{Gomsh}
is the appropriate constant for the bottomonium mesons.
Since the b-quark is heavier than the c-quark, the peaks of
the fragmentation functions shift significantly toward higher values of z.\\
As we explained, the fragmentation is related to the low-energy part of the hadron production
so, in principle, should be extracted from experimental data. However, in practice, it is hard to extract FFs phenomenologically,
especially for the heavy quarkonia   that, at the present, there are no experimental data for them. 
A comprehensive review on the behavior of the gluon FFs obtained phenomenologically,
shows that for the heavy-light mesons (e.g. $B, D$) the peaks of
the gluon FFs occur at low values of the fragmentation parameter $z$ and the FFs approach to zero when either $z\rightarrow 0$ or $z\rightarrow 1$.
For example, the behavior of the $g\rightarrow B$ FF is shown in \cite{Kniehl:2011bk} and the $g\rightarrow D^+$ one is shown in \cite{Kneesch:2007ey}.
These behaviors may be expected for the heavy quarkonia, as shown in Figs.~\ref{fig3}-\ref{fig6}.\\
In Ref.~\cite{MoosaviNejad:2011yp}, we pointed out that in top-quark decays $t\rightarrow b+W^+(+g)$, since the observed mesons can be also produced through
a fragmenting gluon, therefore, to obtain the most accurate result for the energy spectrum of the meson, one has to add
the contribution of gluon fragmentation to the b-quark to produce the outgoing meson.
We showed that the $g\rightarrow B/D$ contribution is appreciable only at a low energy of the observed meson so 
the contribution of the gluon leads to an appreciable reduction in decay rate at low values of the meson energy, 
which is in consistency with the peak position of the gluon FF.
However, the contribution of the gluon cannot be discriminated so, this part of calculation is of more theoretical relevance
and there will be no experimental data for the $g\rightarrow B/D$. But the $g\rightarrow B/D/H_c$ FFs are confirmed by the comparison 
between the theoretical predictions and the experimental measurements of the heavy meson cross sections at the LHC.

\section{Conclusion}
\label{sec:three}
The dominant mechanism to produce the heavy quarkonia at  high
transverse momentum is fragmentation; the production of a high energy parton followed by its fragmenting  into 
the heavy quark-antiquark bound states. Beside the phenomenological approaches, there are some
theoretical models to calculate the fragmentation functions analytically.
In the present work,  using the Suzuki's model  we gave out an analytical expression for the initial scale fragmentation functions of gluon  
to split into the spin-singlet and -triplet S-wave charmonium states.
Our results depend on the transverse momentum of the initial gluon whereas 
in other models the integrations over all freedom degrees are performed.
Since the transverse momentum dependent FFs show up explicitly in several semi-inclusive cross sections, therefore
in the QCD corrections the inclusion of these dependent FFs will be necessary.\\
We discussed that there are three different scenarios to impose the spin effects of heavy quarkonia into the gluon FFs and
we presented the fragmentation functions in each three scenarios. The results show that  the fragmentation 
probability of the gluon into the spin-triplet bound-state  is the biggest one.
However, at present there are no experimental data for the heavy quarkonium 
fragmentation  but a comparison between the treatment of the gluon FFs in Figs.~\ref{fig3}, \ref{fig5} and \ref{fig6} with
the $g\rightarrow B$ \cite{Kniehl:2011bk} and $g\rightarrow D$ \cite{Kneesch:2007ey} FFs can confirm the correctness of 
our derived results.
Our results can be directly used for the S-wave bottomonium sates ($b\bar{b}$-system), with some simple replacements.

\begin{acknowledgments}
We would like to thank the CERN
TH-PH division for its hospitality, where a portion of this
work was performed.
\end{acknowledgments}

\end{document}